\documentclass[submission,copyright,creativecommons]{eptcs}

\newcommand*{\GandALF}{}

\providecommand{\event}{GandALF 2018} % Name of the event you are submitting to
\usepackage{underscore}           % Only needed if you use pdflatex.

\input{Styles/MainStyle.sty}
\input{Styles/UppaalTools.sty}
\input{Styles/Ecdar.sty}
\input{Styles/Glossaries.sty}
\input{Styles/Commands.sty}

\title{Effortless Fault Localisation:\\Conformance Testing of Real-Time Systems in \ecdar}
\def\titlerunning{Effortless Fault Localisation: Conformance Testing of Real-Time Systems in \ecdar}
\ifdefined\GandALF
    \author{
        Tobias R. Gundersen \qquad Florian Lorber \qquad Ulrik Nyman \qquad Christian Ovesen
        \institute{\hspace{.5em} Aalborg University, Denmark \hspace{10em} Aalborg University, Denmark}
        \email{
            gundersen.tobias@gmail.com \hspace{0.2em} 
            florber@cs.aau.dk \hspace{1.0em} 
            ulrik@cs.aau.dk \hspace{1.0em}
            ovesen.chr@gmail.com \hspace{0.1em}
        }
    }
\else
    \author{
        Tobias R. Gundersen \qquad\qquad Christian Ovesen \hspace{0.5em} 
        \institute{Aalborg University, Denmark}
        \email{tgunde13@student.aau.dk \quad\qquad covese13@student.aau.dk}
    }
\fi

\ifdefined\GandALF
    \def\authorrunning{T.R. Gundersen, F. Lorber, U. Nyman \& C. Ovesen}
\else
    \def\authorrunning{T.R. Gundersen \& C. Ovesen}
\fi

\begin{document}
\maketitle

\begin{abstract}
Model checking of real-time systems has evolved throughout the years. Recently, the model checker \ecdar, using timed I/O automata, was used to perform compositional verification. However, in order to fully integrate model checking of real-time systems into industrial development, we need a productive and reliable way to test if such a system conforms to its corresponding model. Hence, we present an extension of \ecdar that integrates conformance testing into a new IDE that now features modelling, verification, and testing. The new tool uses model-based mutation testing, requiring only the model and the system under test, to locate faults and to prove the absence of certain types of faults. It supports testing using either real-time or simulated time. It parallelises test-case generation and test execution to provide a significant speed-up. We also introduce new mutation operators that improve the ability to detect and locate faults. Finally, we conduct a case study with 140 faulty systems, where \pname detects all faults.
\end{abstract}

\section{Introduction}
It is essential to ensure the correctness of safety-critical systems. Model checking can prove the correctness of a design.
However, a correct design does not ensure the correctness of its implementation.
We could test this correctness by writing unit tests by hand.
However, this is time-consuming and error-prone, as we would manually derive them from the requirements of the \gls{SUT}. 

% Model-based is good
Instead, we can test if the \gls{SUT} conforms to the model proven to be correct.
Model checkers for real-time systems focus on timing aspects, which are essential for safety-critical real-time systems.
When testing based on a timed model, the tests will focus on timed aspects as well.

One model-based testing technique is \gls{MBMT}. It has already been applied to \glspl{TA} with simulated time \cite{timeForMutants,MutationGenerationEcdar}. Larsen et al. \cite{MutationGenerationEcdar} use the model checker \ecdar \cite{David2010EcdarTool} to perform unbounded conformance checks, to provide a significant speed-up compared to \cite{timeForMutants}, and to enable adaptive test-cases, which produce fewer inconclusive test verdicts.

% Contributions
The main contributions of this paper are (1) an extension of \ecdar that integrates conformance testing of real-time systems, using only the model and the \gls{SUT}, into the tool; (2) automatic fault localisation through the introduction of primarily failed test-cases; and (3) support for testing using either real-time or simulated time.
%  in order to improve productiveness

Other contributions are parallelisation of test-case generation and test execution to provide a significant speed-up and the introduction of new mutation operators that improve the ability to detect and locate faults.

%Motivation
The rest of the paper is structured as follows: First, in \cref{sec:related} we discuss related work. In \cref{sec:preliminaries} we present preliminaries, including the definition of \glspl{TIOA} and the workflow of \gls{MBMT}. In \cref{sec:integration} we discuss the integration of \gls{MBMT} into \pname. In \cref{sec:caseStudy} we present a case study of our extension. In \cref{sec:ui} we showcase the extension.
%In \cref{sec:discussion} we discuss it with respect to the case study. 
Finally, in \cref{sec:conclusion} we conclude the paper and outline ideas for future work. 
\section{Related Work} \label{sec:related}

\gls{MBMT} has already been applied to probabilistic finite state machines \cite{Hierons2007FiniteStateMachineMBMT} and UML state machines \cite{Aichernig2015}. These approaches mainly focus on testing functional behaviour. Aichernig \etal \cite{timeForMutants} propose to use \gls{MBMT} for \glspl{TA} and presents the tool MoMuT::TA\footnote{\url{https://momut.org/}} that implements test-case generation.

Larsen \etal \cite{MutationGenerationEcdar} propose to use \ecdar to perform unbounded conformance checks, to provide a significant speed-up compared to \cite{timeForMutants}, and to enable adaptive test-cases, which produce fewer inconclusive test verdicts.

Lorber \etal \cite{Lorber2018viaModelChecking} propose an approach to combine \gls{MBMT} and \gls{TCTL} properties used for verification; from a set of generated mutants, they check if the mutants satisfy the properties. For each violation, they use the counterexample as a test-case. This way, the approach only model checks individual properties rather than performing a potentially more time-consuming full conformance check. Also, the approach generates fewer test-cases, which takes less time to execute, and the test-cases focus on the safety-critical properties derived from the requirements of the system.

Devroey \etal \cite{Devroey2016FeaturedMutantModel} propose another way to reduce generation and testing time. They propose to encode each mutant as a product in a software product line; instead of generating individual mutants, they propose to generate a single featured mutant model that -- when configured -- can represent any mutant. This way, the mutants can share execution, allowing for testing in a single run. While the approach is for assessing the quality of an existing test suite, it could also be applied for test-case generation.

There exist tools for model-based testing using \glspl{TA}. The tool \uppaalTron \cite{Hessel2008TronTool} can conduct conformance tests on systems in real-time and thus can be applied to physical systems. It uses an environment \gls{TA} to decide which inputs and delays to trigger and detects whether the \gls{SUT} produces allowed outputs. A tester can easily construct a simple, permissive environment, but at the cost of simply triggering random inputs and delays. Otherwise, they can construct environment models that steer the execution towards critical areas.

The tool \uppaalCover \cite{hessel2007cover} generates test-cases based on \gls{TA} and \gls{TCTL}. A tester can reuse the \gls{TCTL} properties used for verifying the test model. The tester also needs to specify monitoring automata that describe coverage criteria. 

The tool \uppaalYggdrasil \cite{Kim2015} is similar to \uppaalCover. It generates test-cases based on \gls{TA}, \gls{TCTL} properties, and random search. 

Contrary to these three tools, the presented \ecdar extension uses a fault-based approach that can prove the absence of certain types of faults. Furthermore, it generates test-cases only based on a test model. Thus, an \ecdar tester need only provide the test model and the \gls{SUT}; they do not need to perform the time-consuming and error-prone tasks of constructing environment models, \gls{TCTL} properties, or monitoring automata.

\section{Preliminaries}\label{sec:preliminaries}
In this section we define \glspl{TIOA}, their underlying transition systems, determinism, input-enabledness, and refinement. Finally, we describe the workflow of \gls{MBMT}.

\clearpage
\subsection{Timed I/O Automata}

A \glspl{TIOA} \cite{TIOA} is a syntactical, finite representation of a timed system. It is a tuple $(\mcal{Q}, q_0, \mcal{C}, \mcal{V}, \Sigma, \mcal{E}, \mcal{I})$, where:
\begin{itemize}[noitemsep]
    \item $\mcal{Q}$ is a finite set of locations.
    
    \item $q_0 \in \mcal{Q}$ is the initial location.
    
    \item $\mcal{C}$ is a finite set of clocks used to represent time.
    
    \item $\mcal{V}$ is a finite set of integer variables local to the automata. Each variable $v \in \mcal{V}$ has lower and upper bounds $v_{min}, v_{max} \in \mathbb{Z}$ and an initial value $v_0 \in [v_{min}, v_{max}]$.
    
    \item $\Sigma$ is a finite set of observable actions partitioned into inputs ($\Sigma_i$) and outputs ($\Sigma_o$).
    
    \item $\mcal{E}$ is a finite set of edges of the form $e=(q, g, \sigma, R, u, q')$, where:
    \begin{itemize}[noitemsep]
        \item $q, q' \in \mcal{Q}$ are the source and target locations, respectively.
        \item $g$ (the \emph{guard}) is a conjunction $\bigwedge_{b \in B_e} b$, \ie $B_e$ is the set of basic constraints of the guard of edge $e$. Each basic constraint $b$ is of the form $x \circ c$, where $x \in \mcal{C} \cup \mcal{V}$, $\circ \in \{<, \leq, =, \neq, \geq, >\}$, and $c \in \mathbb{Z}$. The guard must be satisfied when executing the edge.
        \item $\sigma \in \Sigma$ is the observable action.
        \item $R \subseteq \mcal{C}$ is the set of clocks to reset.
        \item $u: \mcal{V} \to \mathbb{Z}$ is an update of some of the variables. For each such variable $v$, $u(v) \in [v_{min}, v_{max}]$.
    \end{itemize}
    
    \item $\mcal{I}: \mcal{Q} \to \mcal{U}(\mcal{C})$ is a set of invariants for some of the locations. We write $\mcal{U}(\mcal{C})$ for the set of constraints over $\mcal{C}$ of the form $x \circ c$, where $x \in \mcal{C}$, $\circ \in \{<, \leq\}$, and $c \in \mathbb{N}$. An invariant must be satisfied when entering and while in the respective location.
\end{itemize}

\begin{figure}[t]
    \centering
    \includegraphics[width=.6\textwidth]{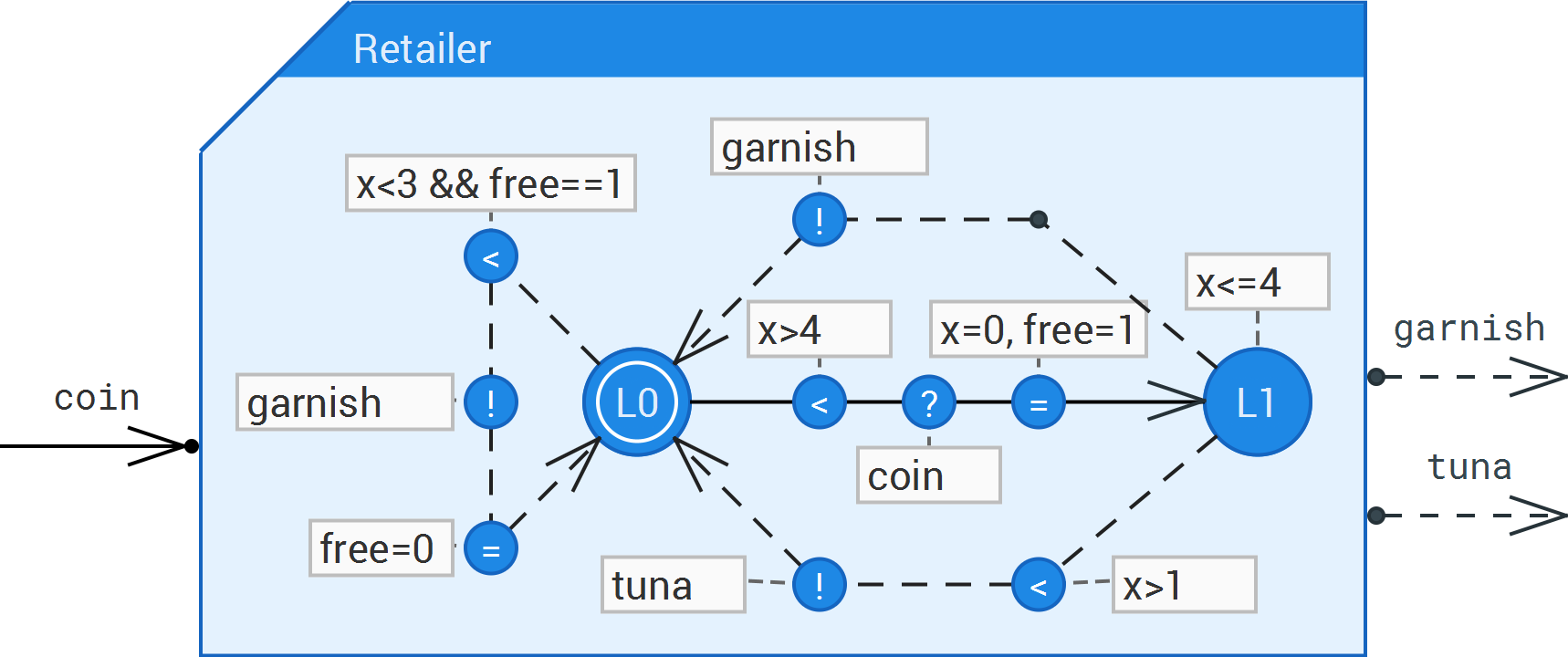}
    \caption{An \pname \gls{TIOA} \eComp{Retailer} for a fish retailer.}
    \label{fig:tioaExample}
\end{figure}

Contrary to \cite{TIOA}, our definition of \gls{TIOA} includes local variables, since the \ecdar engine supports them.
\cref{fig:tioaExample} illustrates the \gls{TIOA} for which
$\mcal{Q} = \{L0, L1\}$,
$q_0 = L0$, 
$\mcal{C} = \{x\}$, 
$\mcal{V} = \{\mathit{free}\}$, $\mathit{free}_0=\mathit{free}_{min}$ $=0$, $\mathit{free}_{max}=1$,
$\Sigma_i = \{coin?\}$,
$\Sigma_o = \{garnish!, tuna!\}$,
$\mcal{E} = \{(L0, x<3 \land free=1, garnish!,$ $\emptyset,$ $\{(free, 0)\},$ $L0)$,
$(L0, x>4, coin?, \{x\}, \{(free, 1)\}, L1),$
$(L1, true, garnish!, \emptyset, \emptyset, L0),$
$(L1, x>1, tuna!, \emptyset, \emptyset, L0)\}$, and
$\mcal{I} = \{(L1, x\leq4)\}$.

We denote by $\mcal{A}$ the set of all \glspl{TIOA}. 
We denote by $\mcal{E}_i$ the input edges $\{e = (q, g, \sigma, R, u, q') \mid e \in \mcal{E} \land \sigma \in \Sigma_i\}$, and by $\mcal{E}_o$ the output edges $\{e = (q, g, \sigma, R, u, q') \mid e \in \mcal{E} \land \sigma \in \Sigma_o\}$.

%TIOTS
\subsection{Timed I/O Transition Systems}

A \gls{TIOTS} \cite{TIOA} $S$ is the semantic representation induced by a \gls{TIOA} $A$, written $S = [\![A]\!]_{sem}$. It is a tuple $(St, s_0, \Sigma, \rightarrow)$ where:

\begin{itemize}[noitemsep]
    \item $St$ is a set of states.
    \item $s_0 \in St$ is the initial state.
    \item $\Sigma$ is a finite set of observable actions partitioned into inputs ($\Sigma_i$) and outputs ($\Sigma_o$).
    \item $\rightarrow \subseteq St \times (\Sigma \cup \mathbb{R}_{\geq0}) \times St$ is a transition relation.
\end{itemize}

We write $s \xrightarrow{\smash{a}} s'$ instead of $(s, a, s') \in \rightarrow$.
We write $s \xrightarrow{\smash{a}}$ for $\exists s' \ldotp s \xrightarrow{\smash{a}} s'$.
We use $i?$, $o!$, and $d$ to range over inputs, outputs, and delays ($\mathbb{R}_{\geq0}$), respectively.

\paragraph{Determinism:}
A \gls{TIOTS} is \emph{deterministic} iff $\forall s, s', s'' \in St \ldotp \forall \sigma \in \Sigma \ldotp s \xrightarrow{\smash{\sigma}} s' \land s \xrightarrow{\smash{\sigma}} s'' \implies s' = s''$. That is, whenever the transition system can perform an action, there is always only one transition to take with that action. As an example, the transition system of $Retailer$ in \cref{fig:tioaExample} is deterministic.

\paragraph{Input-Enableness:}
A \gls{TIOTS} is \emph{input-enabled}, iff $\forall s \in St \ldotp \forall i? \in \Sigma_i \ldotp s \xrightarrow{\smash{i?}}$. That is, it can always accept any of its defined inputs. As an example, the transition system of $Retailer$ in \cref{fig:tioaExample} does not accept a \eSync{coin?} in \eLoc{L1}, and thus it is not input-enabled.

% Angelic, demonic, universal
We can use \emph{angelic completion} \cite{angelic2008} to transform an automaton to one with an input-enabled transition system: For each input, it receives new self-loops for each state that did not accept that input. It corresponds to ignoring those inputs.

Alternatively, we can use \emph{demonic completion} \cite{demonic2004} to transform an automaton to one with an input-enabled transition system: Instead of self-loops, it has new edges leading to a \emph{universal location}. In a universal location, every possible behaviour defined by the automata is enabled. That is, it can continuously both delay indefinitely and perform every action $\Sigma$.

\subsection{Refinement}
Refinement \cite{TIOA} compares the behaviour of two deterministic, input-enabled transition systems.
A \gls{TIOTS} $T = (St^T, t_0,\Sigma,\rightarrow^T)$ \emph{refines} a \gls{TIOTS} $S = (St^S, s_0, \Sigma, \rightarrow^S)$, written $T \leq S$, iff there exists a binary relation $R \subseteq St^T \times St^S$ containing $(t_0,s_0)$ such that for each pair of states $(t,s) \in R$ we have:
\begin{itemize}[noitemsep]
    \item $\forall s' \in St^S \ldotp \forall i? \in \Sigma_i \ldotp s \xrightarrow{\smash{i?}}^S s' \implies \exists t' \in St^T \ldotp t \xrightarrow{\smash{i?}}^T t' \land (t',s') \in R$.
    \item $\forall t' \in St^T \ldotp \forall o! \in \Sigma_o \ldotp t \xrightarrow{\smash{o!}}^T t' \implies \exists s' \in St^S \ldotp s \xrightarrow{\smash{o!}}^S s' \land (t',s') \in R$.
    \item $\forall t' \in St^T \ldotp \forall d \in \mathbb{R}_{\geq0} \ldotp t \xrightarrow{\smash{d}}^T t' \implies \exists s' \in St^S \ldotp s \xrightarrow{\smash{d}}^T s' \land (t',s') \in R$.
\end{itemize}

$T \leq S$ represents that $T$ has less behaviour than or equal behaviour to $S$. A \gls{TIOA} $A_1$ refines another \gls{TIOA} $A_2$, written $A_1 \leq A_2$, iff $[\![A_1]\!]_{sem} \leq [\![A_2]\!]_{sem}$. That is, there is a corresponding refinement between their underlying \glspl{TIOTS}.

\subsection{Model-Based Mutation Testing} \label{sec:modelBasedMutationTesting}

\begin{figure}[t]
    \centering
    \begin{subfigure}[b]{0.5\textwidth}
        \centering\includegraphics[width=.95\textwidth]{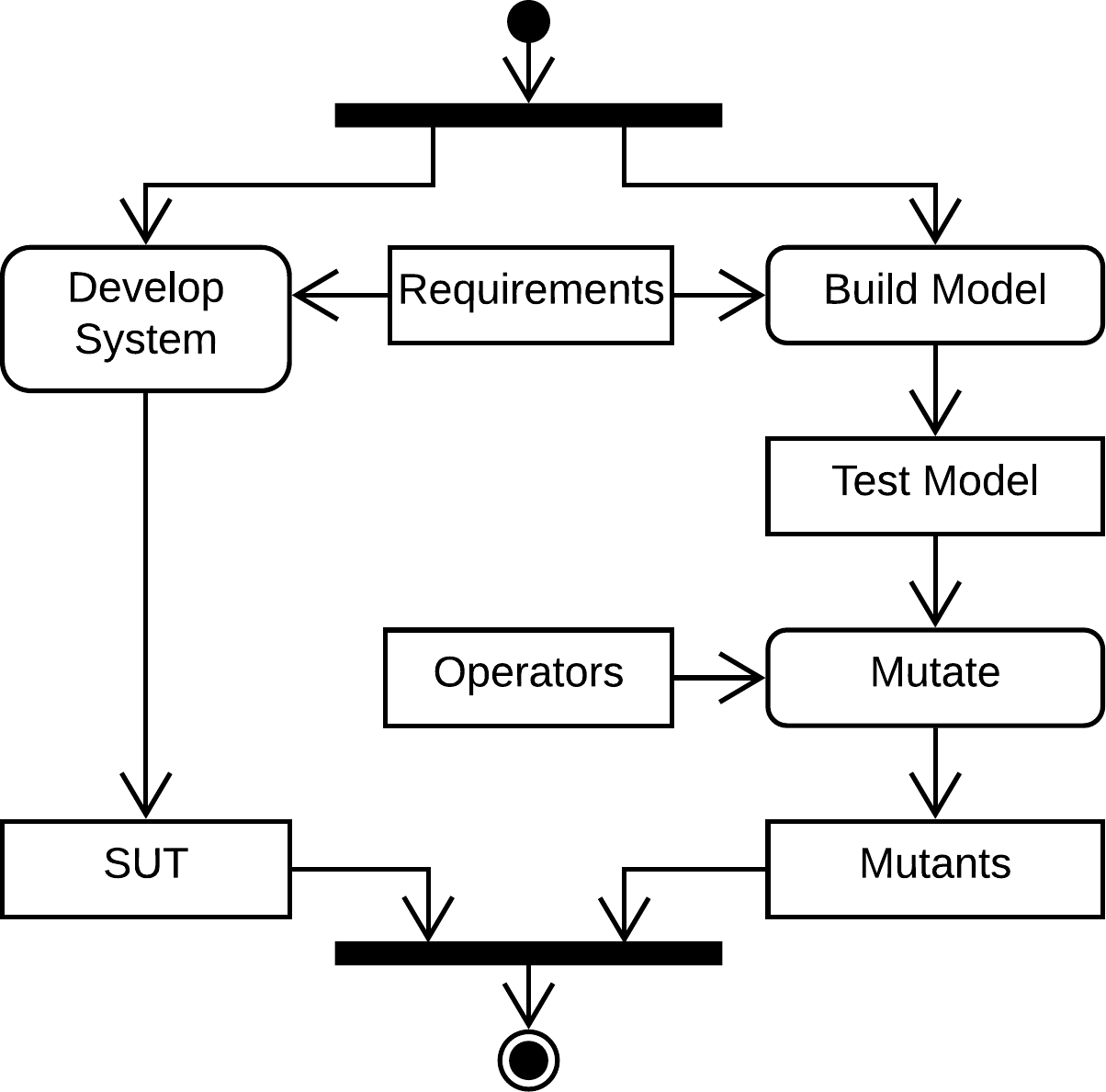}
        \caption{\label{fig:mutationTesting1}Activity to create an \gls{SUT}, a test model, and mutants.}
    \end{subfigure}%
    \begin{subfigure}[b]{0.5\textwidth}
        \centering\includegraphics[width=.95\textwidth]{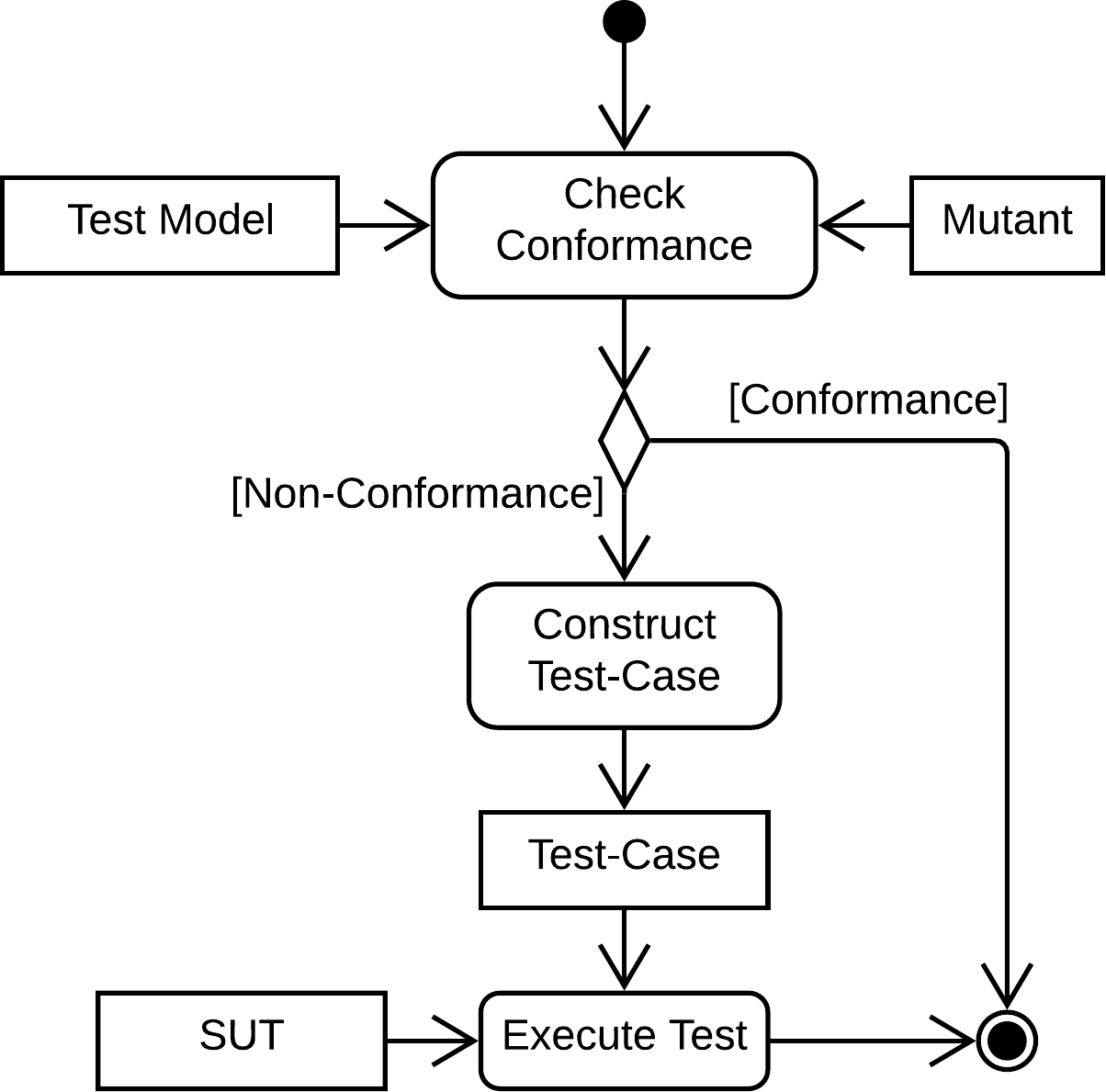}
        \caption{\label{fig:mutationTesting2}Activity to generate and execute a test-case.}
    \end{subfigure}
    \caption{UML 2.5 \cite{uml} activity diagrams of \gls{MBMT}.}
    \label{fig:mutationTesting}
\end{figure}

In \gls{MBMT} we construct a test suite based on mutants of a test model. Firstly, based on requirements of a system, we develop the \gls{SUT} and a test model that it should conform to (see \cref{fig:mutationTesting1}). We then mutate the test model according to some selected mutation operators.
An operator represents certain ways of changing the model, \eg changing the source location of an arbitrary edge to an arbitrary location. A mutant is the result of a single application of an operator and represents a potential fault.

For each mutant, we check if it conforms to the test model (see \cref{fig:mutationTesting2}). If it does, it does not introduce any observable faults. Thus, we discard it. Otherwise, the conformance check provides a counterexample, \ie a way to potentially reveal the fault. We use the counterexample as a test-case by applying it on the \gls{SUT}. A test passes iff the \gls{SUT} is shown to behave according to the test model and not according to the mutant.

Assuming determinism, for all passing test-cases, \gls{MBMT} guarantees that the faults represented by their corresponding mutants do not exist in the \gls{SUT}. A failed test-case is an aid for debugging as presented in  \cite{DebuggingTAMutation}; a failed test-case can show the location and type of the fault and provides a way to reproduce it.

\paragraph{Adaptiveness}
Test-cases can be more or less adaptive. A non-adaptive test-case only covers a trace with its counterexample; if the test model has multiple choices of delaying and outputting, a non-adaptive test-case can only handle one of the choices. If the \gls{SUT} performs an unexpected, allowed delay or output, we assign it the \emph{inconclusive} verdict.

An \emph{adaptive} test-case can handle various choices made by the \gls{SUT} and steer the execution towards the fault represented by the mutant.

\section{Integration}\label{sec:integration}
We divide testing with \pname into three steps: Mutation, test-case generation, and test execution. In this section, we discuss each of these steps and then discuss performance.

\subsection{Mutation} \label{sec:mutation}
A mutation operator is a function $\mcal{M} : \mcal{A} \to 2^\mcal{A}$.
We use the following mutation operators defined by Aichernig \etal \cite{timeForMutants}:

\definecolor{oldOpColour}{HTML}{0000A0}
\newcommand{\opSource}{{\color{oldOpColour}$\mcal{M}_s$}\xspace}
\newcommand{\opTarget}{{\color{oldOpColour}$\mcal{M}_t$}\xspace}
\newcommand{\opInv}{{\color{oldOpColour}$\mcal{M}_{inv}$}\xspace}
\newcommand{\opSink}{{\color{oldOpColour}$\mcal{M}_{sl}$}\xspace}
\newcommand{\opClockReset}{{\color{oldOpColour}$\mcal{M}_{c}$}\xspace}
\newcommand{\opOut}{{\color{oldOpColour}$\mcal{M}_o$}\xspace}
\begin{enumerate}
    \item[\opSource] replaces the \emph{source} location of an edge with another location.
    
    \item[\opTarget] replaces the \emph{target} location of an edge with another location.
    
    \item[\opOut] replaces the action of an edge with a (different) \emph{output}.
    
    \item[\opInv] loosens a constraint in an \emph{invariant} by 1 time unit (\eg $x<=2$ would be loosened to $x<=3$). We do not tighten invariants, as this would result in a conformance.
    
    \item[\opSink] changes the target location of an edge to a new \emph{sink location}. Sink locations accept but ignore all inputs.
    
    \item[\opClockReset] inverts a \emph{clock} reset on an edge; if the clock was originally reset, the reset is removed, otherwise a reset is added.
\end{enumerate}

We define the following new mutation operators. In the definition of the operators, we mutate a \gls{TIOA} $S=(\mcal{Q}, q_0, \mcal{C}, \mcal{V}, \Sigma, \mcal{E}, \mcal{I})$:

\definecolor{newOpColour}{HTML}{A000A0}
\newcommand{\opIn}{{\color{newOpColour}$\mcal{M}_i$}\xspace}
\newcommand{\opInMath}{{\color{newOpColour}\mcal{M}_i}}
\newcommand{\opGuardConstant}{{\color{newOpColour}$\mcal{M}_{gc}$}\xspace}
\newcommand{\opGuardConstantMath}{{\color{newOpColour}\mcal{M}_{gc}}}
\newcommand{\opGuardOpClocks}{{\color{newOpColour}$\mcal{M}_{goc}$}\xspace}
\newcommand{\opGuardOpClocksMath}{{\color{newOpColour}\mcal{M}_{goc}}}
\newcommand{\opGuardOpVars}{{\color{newOpColour}$\mcal{M}_{gov}$}\xspace}
\newcommand{\opGuardOpVarsMath}{{\color{newOpColour}\mcal{M}_{gov}}}
\newcommand{\opVarUpdate}{{\color{newOpColour}$\mcal{M}_{vu}$}\xspace}
\newcommand{\opVarUpdateMath}{{\color{newOpColour}\mcal{M}_{vu}}}

\begin{figure}[tb]
    \centering
    \begin{subfigure}[b]{0.5\textwidth}
        \centering\includegraphics[width=\textwidth]{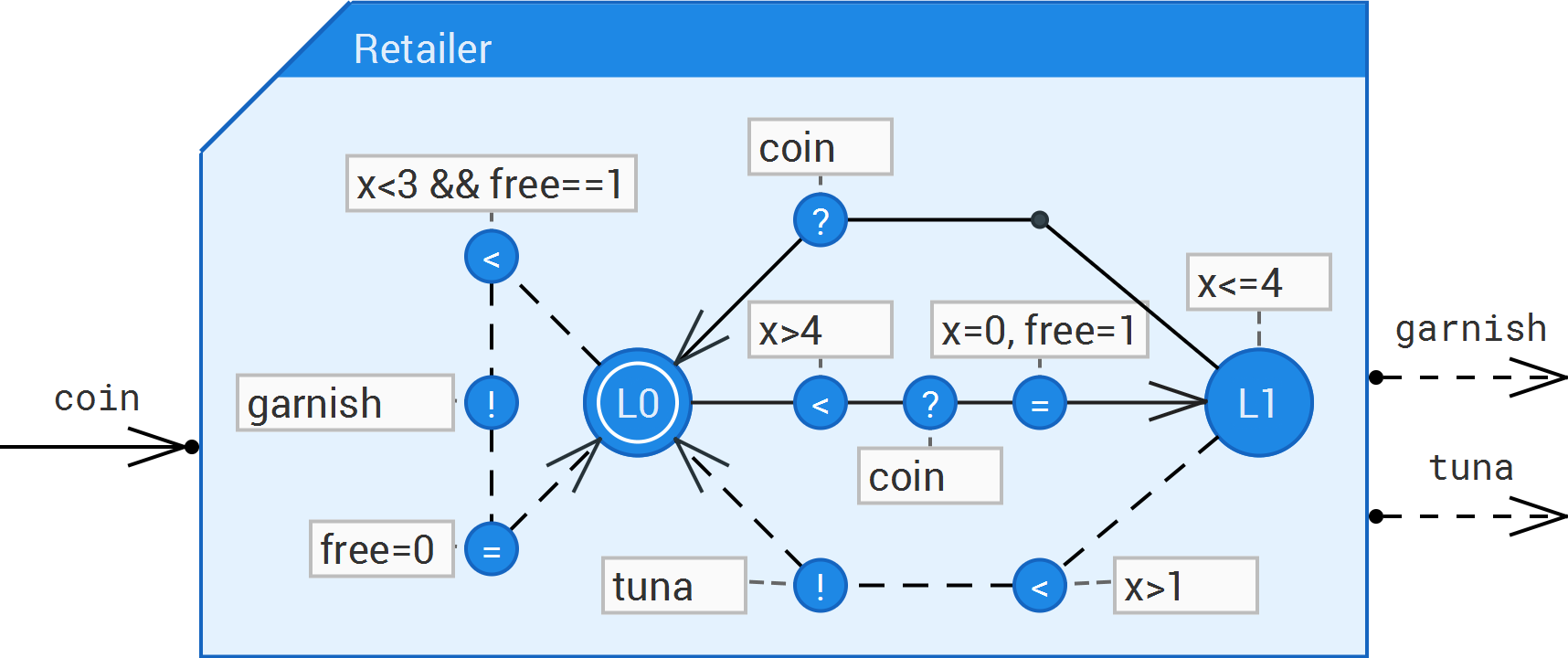}
        \captionsetup{width=.9\textwidth}
        \caption{\label{fig:opInExample}Mutant in $\opInMath(Retailer)$. The action of an edge from \eLoc{L1} to \eLoc{L0} is changed from output \eSync{garnish!} to input \eSync{coin?}.}
    \end{subfigure}%
    \begin{subfigure}[b]{0.5\textwidth}
        \centering\includegraphics[width=\textwidth]{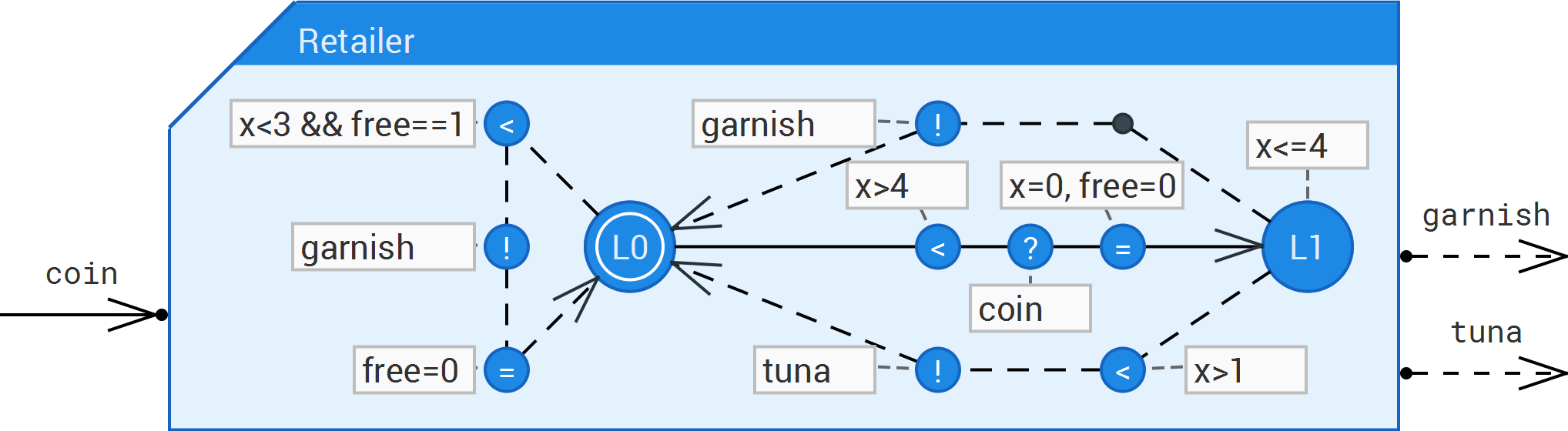}
        \captionsetup{width=.9\textwidth}
        \caption{\label{fig:opVarUpdateExample}Mutant in $\opVarUpdateMath(Retailer)$. The variable update of the edge from \eLoc{L0} to \eLoc{L1} is changed from $(free, 1)$ to $(free, 0)$.}
    \end{subfigure}
    \caption{Two mutants of the \eComp{Retailer} \gls{TIOA} on \cref{fig:tioaExample}.}
    \label{fig:mutExamples}
\end{figure}

\begin{enumerate}
    \item[\opIn] replaces the action of an edge with a (different) \emph{input}.
    This creates $|\mcal{E}_i| (|\Sigma_i|-1) +|\mcal{E}_o| |\Sigma_i|$ mutants.
    Aichernig \etal \cite{timeForMutants} only replaces actions with outputs (the operator \opOut).
    A \gls{TIOA} $M \in \opInMath(S)$ iff $M=(\mcal{Q}, q_0, \mcal{C}, \mcal{V}, \Sigma, (\mcal{E}\setminus\{e_S\}) \cup \{e_M\}, \mcal{I})$, such that $e_S = (q, g, \sigma_S, R, u, q') \in \mcal{E}$, $e_M = (q, g, \sigma_M, R, u, q')$, $\sigma_M \in \Sigma_i$, and $\sigma_S \neq \sigma_M$. An example of such a mutant is given is \cref{fig:opInExample}.
    
    \item[\opGuardConstant] adds or subtracts 1 to or from a \emph{guard constant}. This creates $2 \sum_{e \in \mcal{E}}|B_e|$ mutants. 
    A \gls{TIOA} $M \in \opGuardConstantMath(S)$ iff $M=(\mcal{Q}, q_0, \mcal{C}, \mcal{V}, \Sigma, (\mcal{E}\setminus\{e_S\}) \cup \{e_M\}, \mcal{I})$, such that 
    $e_S = (q, g_S, \sigma, R, u, q') \in \mcal{E}$,
    $e_M = (q, g_M, \sigma, R, u, q')$, 
    $g_S = \bigwedge_{i \in I}(x_i \circ_i c_i^S)$, 
    $g_M = \bigwedge_{i \in I}(x_i \circ_i c_i^M)$, 
    $\exists i' \in I \qdot c_{i'}^S \pm 1 = c_{i'}^M$, 
    and $\forall i \in I \setminus \{i'\} \qdot c_i^S = c_i^M$.
    
    \item[\opGuardOpClocks] changes a \emph{guard operator} with a \emph{clock} as its left side. This creates up to $2 \sum_{e \in \mcal{E}}|B_{e,\mcal{C}}|$ mutants, where $B_{e, \mcal{C}} = \{b \in B_e \mid b = x \circ c \land x \in \mcal{C}\}$. Time is continuous. Thus, in practice we expect that for all clock valuations $v$ we have that $\forall c \in \mathbb{Z} \qdot v < c \iff v \leq c \land v > c \iff v \geq c \land v \neq c $. For this reason, we only mutate with $\leq$ and $>$. With reduced number of guard operators, we reduce generation and test execution time. \opGuardOpClocks overlaps with two mutation operators in \cite{timeForMutants}: $\mu_{cg}$ that in a single mutation changes all operators in a guard to one among $\{<, \leq, =, \geq, >\}$, and $\mu_{ng}$ that negates a guard. Since a mutation is a simple syntactic change \cite{analysisAndSurveyMutation}, we combine these two operators into \opGuardOpClocks that only changes a single operator. 
    A \gls{TIOA} $M \in \opGuardOpClocksMath(S)$ iff $M=(\mcal{Q}, q_0, \mcal{C}, \mcal{V}, \Sigma, 
    (\mcal{E}\setminus\{e_S\}) \cup \{e_M\}, \mcal{I})$, such that 
    $e_S = (q, g_S, \sigma, R, u, q') \in \mcal{E}$,
    $e_M = (q, g_M, \sigma, R, u, q')$, 
    $g_S =  \bigwedge_{i \in I}(x_i \circ_i^S c_i)$, 
    $g_M = \bigwedge_{i \in I}(x_i \circ_i^M c_i)$, 
    $\exists i' \in I \qdot x_{i'} \in \mcal{C} \land \circ_{i'}^M \in \{\leq, >\} \setminus \{\circ_{i'}^S\}$, 
    and $\forall i \in I \setminus \{i'\} \qdot \circ_i^S = \circ_i^M$.
    
    \item[\opGuardOpVars] changes a \emph{guard operator} with a \emph{variable} as its left side. This creates $5 \sum_{e \in \mcal{E}}|B_{e,\mcal{V}}|$ mutants, where $B_{e, \mcal{V}} = \{b \in B_e \mid b = x \circ c \land x \in \mcal{V}\}$. 
    A \gls{TIOA} $M \in \opGuardOpVarsMath(S)$ iff $M=(\mcal{Q}, q_0, \mcal{C}, \mcal{V}, \Sigma, (\mcal{E}\setminus\{e_S\}) \cup \{e_M\}, \mcal{I})$, such that 
    $e_S = (q, g_S, \sigma, R, u, q') \in \mcal{E}$,
    $e_M = (q, g_M, \sigma, R, u, q')$, 
    $g_S = \bigwedge_{i \in I}(x_i \circ_i^S c_i)$, 
    $g_M = \bigwedge_{i \in I}(x_i \circ_i^M c_i)$, 
    $\exists i' \in I \qdot x_{i'} \in \mcal{V} \land \circ_{i'}^M \in \{<, \leq, =, \neq, \geq, >\} \setminus \circ_{i'}^S$, 
    and $\forall i \in I \setminus \{i'\} \qdot \circ_i^S = \circ_i^M$.
    
    \item[\opVarUpdate] assigns a value to a local \emph{variable} in an \emph{update} property. If the variable is already being assigned in this property, the mutating assignment overrides the existing one. If the existing and mutating assignment values are equal, the corresponding mutant is not created. This creates up to $|\mcal{E}| \sum_{v \in \mcal{V}}(v_{max}-v_{min}+1)$ mutants. 
    A \gls{TIOA} $M \in \opVarUpdateMath(S)$ iff $M=(\mcal{Q}, q_0, \mcal{C}, \mcal{V}, \Sigma, (\mcal{E}\setminus\{e_S\}) \cup \{e_M\}, \mcal{I})$, such that
    $e_S = (q, g, \sigma, R, u_S, q') \in \mcal{E}$, 
    $e_M = (q, g, \sigma, R, u_M, q')$, 
    $\exists v' \in \mcal{V} \qdot u_M(v') \neq u_S(v') \lor (v' \notin \text{dom}(u_S) \land v' \in \text{dom}(u_M))$, and 
    $\forall v \in \mcal{V} \setminus \{v'\} \qdot u_M(v) = u_S(v) \lor (v \notin \text{dom}(u_S) \land v \notin \text{dom}(u_M))$. 
    An example of such a mutant is given is \cref{fig:opVarUpdateExample}.
\end{enumerate}

The definition of mutation operators is simplified compared to the \pname implementation for better understanding.
For instance, the implementation can also handle more complex constraints, \eg constraints with addition, subtraction, and mixed use of constants, clocks, and variables on either side.
Also, it does not mutate certain locations and edges. For instance, it is sometimes inappropriate to mutate the universal location and its outgoing edges.

\subsection{Test-Case Generation} \label{sec:generation}
% Discard non-deterministic models
We use the approach presented in \cite{MutationGenerationEcdar} to perform conformance checks. That is, we use the \ecdar engine to determine if the mutant refines the test model. Since refinement assumes determinism, we discard non-deterministic models.

% we should not discard non-input-enabled models
Refinement also assumes input-enabledness. However, we do not want to force the modeller to make the test model input-enabled. Rather, the behaviour missing in order to be input-enabled is not relevant for the \gls{SUT} and thus should not be tested for.

% Instead, we use angelic/demonic completions
Instead, we apply demonic completion on the test model and angelic completion on the mutants like the approach presented in \cite{MutationGenerationEcdar}. This way, traces leading to missing behaviour will transition the test model into the universal location. Everything refines the universal location. Thus, mutants resulting in such traces will not yield a counterexample for the refinement.

The \ecdar engine solves a refinement check as a timed game. To check if $T \leq S$, the goal of the game is to find a strategy for revealing the non-refinement $T \not\leq S$ by triggering delays and inputs, and observing outputs. In the case of a non-refinement, the \ecdar engine produces a strategy. A strategy can handle various choices made by the \gls{SUT} and can steer the execution towards the goal. Thus, it provides us with an adaptive test-case.

\subsection{Test Execution} \label{sec:testExecution}
We develop a test driver that executes the generated test-cases on an \gls{SUT}. The test driver implementation is inspired by the algorithm presented by Larsen et al. \cite{MutationGenerationEcdar}.

%Describe how the test driver communicates with the SUT
The test driver communicates with the \gls{SUT} over its standard I/O streams. We treat the \gls{SUT} as a black box where the inputs and outputs we send and receive are the same as those represented in the test model. The driver can test systems using either real-time or simulated time. Using real-time, we can test physical systems. However, using real-time is often significantly slower because it may need to perform physical delays.
When simulating time, rather than physically delaying the test, the test driver computes and sends to the \gls{SUT} how long the \gls{SUT} is allowed to simulate delay without interruptions through an input, and the \gls{SUT} answers how long it actually simulated.

If the \gls{SUT} terminates while testing, we treat it as a sink location. This allows us to finish a test if the program terminates.

\paragraph{Rules}
%Define strategy and what the test driver does with it
A strategy consists of delay, input, and output rules, all with disjoint conditions. The test driver checks which rule in the strategy is satisfied for the current states of the test model and the mutant.

If the satisfied rule is an \emph{input rule}, we send an input to the \gls{SUT}. For a \emph{delay rule} the test driver performs a delay until the rule is no longer satisfied or the \gls{SUT} has produced an output. The test driver cannot force the \gls{SUT} to perform an output. It instead must wait for the \gls{SUT} to produce one. Thus, we treat an \emph{output rule} as a delay rule.

\paragraph{Aborting}

A location with no invariant but with outgoing output edges with fully permissive guards implies that the \gls{SUT} can wait forever before outputting. This behaviour causes output rules to suggest that we wait indefinitely until the \gls{SUT} (hopefully) outputs. To avoid this, we introduce a maximum wait time; if this is exceeded, we abort the current test.

%Finishing remarks on and definition of steps.
If the \gls{SUT} avoids the states needed to determine the existence of a fault by looping among the same set of states, the strategy will suggest that we loop indefinitely among the same set of rules. To avoid this, we enforce a bound. Whenever we change the current rule, we increment a step value. If the value exceeds the bound, we abort the current test. 

\paragraph{Verdicts}

\definecolor{priFailColour}{HTML}{00A0A0}
\definecolor{dataPriFailColour}{HTML}{00D0D0}
\definecolor{passColour}{HTML}{00A000}
\definecolor{norFailColour}{HTML}{A00000}
\definecolor{dataNorFailColour}{HTML}{D00000}

\begin{table}[tb]
    \centering
    \begin{tabular}{llll}%
         & \multicolumn{3}{c}{Test model} \\
         & \multicolumn{1}{l|}{} & \multicolumn{1}{c|}{\cmark} & \multicolumn{1}{c}{\xmark} \\ \cline{2-4} 
        \multirow{2}{*}{Mutant} & \multicolumn{1}{l|}{\cmark} & \multicolumn{1}{l|}{Continue} & \color{priFailColour}Primary fail \\ \cline{2-4} 
         & \multicolumn{1}{l|}{\xmark} & \multicolumn{1}{l|}{\color{passColour}Pass} & \color{norFailColour}Other fail
    \end{tabular}
    \caption{What action to take based on whether the test model and the mutant can (\cmark)  or cannot (\xmark) simulate a delay or an output produced by the \gls{SUT}.}
    \label{fig:verdicts}
\end{table}

%Handling of test results
As described in \cref{sec:modelBasedMutationTesting} a test can pass, fail or be inconclusive. We simulate the actions of the \gls{SUT} on the test model and the mutant to determine if the current test passes or fails according to \cref{fig:verdicts}. A test {\color{passColour}passes} iff the test model can simulate a delay or an output, but the mutant cannot.

A test {\color{norFailColour}fails} iff the test model cannot simulate a delay or an output. If a test fails, we simulate the failing action on the mutant. If the mutant can perform it, then we have found a fault that is recognised by the mutant. This mutant is especially helpful for locating the fault. We call these types of fails {\color{priFailColour}primary fails}.
 
A test is inconclusive, if there are no applicable rules for the current states of the models or if we abort the test.

\subsection{Performance}
% Generation in concurrent threads
Generation is a computationally heavy task for which the CPU is the bottleneck. However, we call the \ecdar engine through its command-line interface, which causes delays, making the CPU underutilised. To speed up generation, we generate multiple test-cases in parallel, making \ecdar able to fully utilise the CPU.

% pipeline
Executing a test is not necessarily a computationally heavy task for the test device. Thus, we could speed up testing by generating and executing tests in a pipeline manner. However, we want a consistent and realistic test environment. Running computationally heavy tasks in the background while testing will slow down the test driver, which can cause false positive verdicts (\eg if the \gls{SUT} would have outputted too early). Thus, we test only after we have generated all test-cases.

% Multiple SUT instances
Some systems allow running multiple concurrent instances of them without problems. For this reason, \ecdar allows running multiple concurrent instances of the \gls{SUT}, which speeds up test execution. This is especially useful for real-time testing, as the test driver might wait a long time for the \gls{SUT} to output. By default, we run only one concurrent instance. We leave it up to the user to set a low enough limit on the maximum number of concurrent instances.

\section{Case Study} \label{sec:caseStudy}

\begin{figure}[tb]
    \centering
    \includegraphics[width=\textwidth]{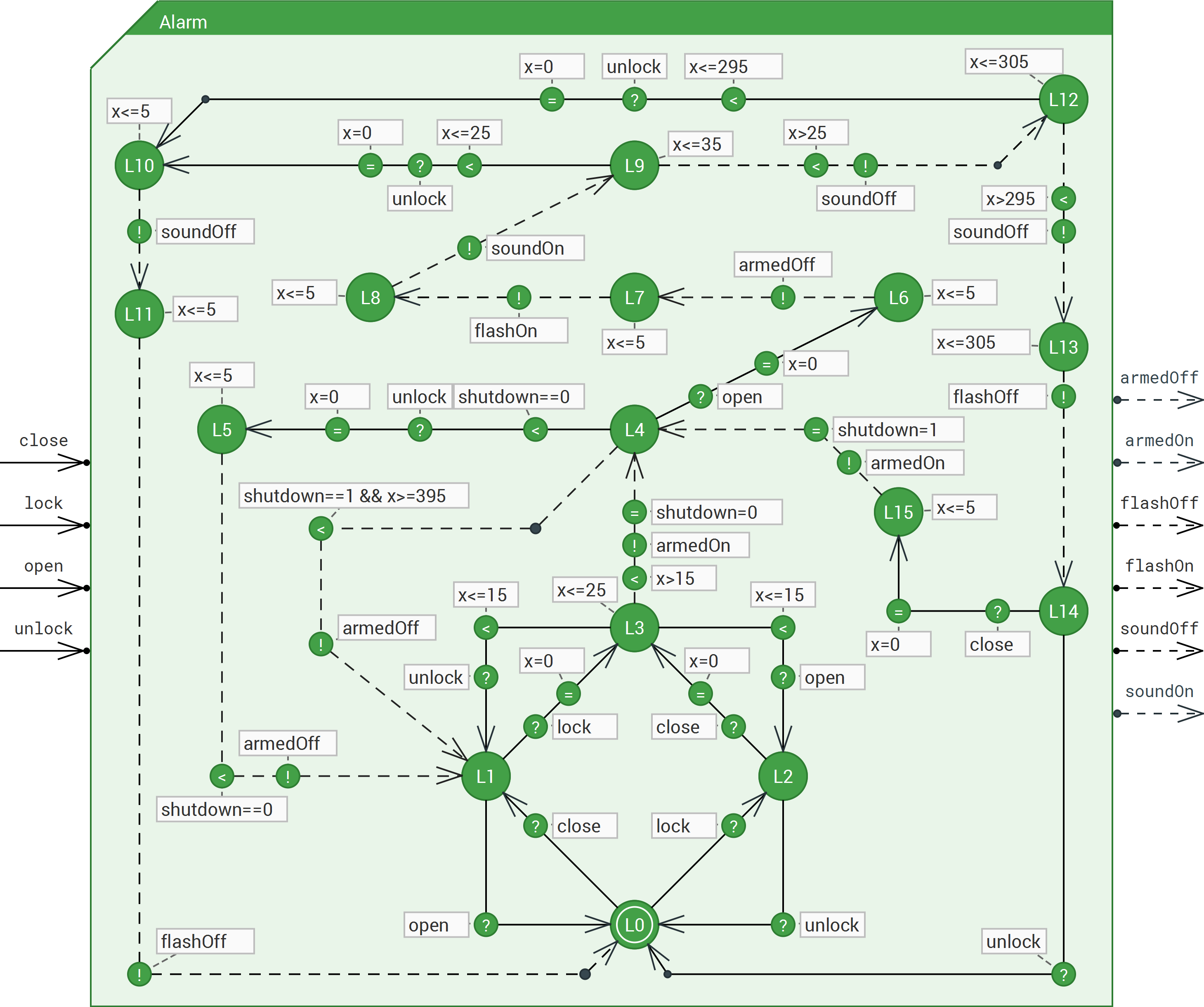}
    \caption{An \pname \gls{TIOA} \eComp{Alarm} for a car alarm system.}
    \label{fig:alarm}
\end{figure}

We implement a modified version of the car alarm system from \cite{MutationGenerationEcdar} to be used as a case study for the purpose of evaluating \pname. The system represents a car alarm that is armed and triggered if someone opens a door without unlocking the car first. If triggered, it can be disarmed or -- if left alone -- it disarms itself after a set duration. As we have defined mutation operators that handle variables, we add a \eVar{shutdown} variable. Furthermore, we make the model robust with regards to time constraints in order to support testing using real-time. The model of the modified system can be seen on \cref{fig:alarm}.

\pname detected faults in our initial implementation. We used \pname to locate and fix the faults. For this case study, we use the final version of the implementation. It has no faults according to \pname.

We perform our tests on a Windows 10 Pro v. 1803 computer with an AMD Ryzen 7 1690 CPU and two Kingston HyperX SH103S3/120G SSDs in Raid 0. We mutate using all \pname mutation operators. We define 1 time unit as 1 s, the number of concurrent \gls{SUT} instances as 5, the maximum wait time as 420 time units, and the step bound as 40 (see \cref{sec:ui}).

It took 0.61 s to generate all 1173 mutants and 30 s to generate all 772 test-cases. A test execution using real-time took 3 h 52 m, while an execution using simulated time took 26 s, which is a significant speed-up.

To evaluate our integration of \pname we use the mutation testing tool PIT\footnote{\url{http://pitest.org/} - version 1.3.2} to generate variations of the car alarm system. We used all mutation operators from PIT and mutated the \texttt{CarAlarm} class (see \cref{sec:ui}) to generate 278 systems.

\begin{table}[tb]
    \newcommand{\f}{{%
        \cellcolor{dataNorFailColour}%
    }}
    \newcommand{\p}{{%
        \cellcolor{dataPriFailColour}%
    }}
    
    \setlength{\tabcolsep}{0pt} % Removes the default of 6pt hskip between columns
    \newcommand{\defaultHskip}{\hskip 6pt}
    \newcommand{\cellSize}{0.97mm}

    \centering
    \begin{tabular}{|@{\defaultHskip}l@{\defaultHskip}|@{\defaultHskip}r@{\defaultHskip}|*{25}{p{\cellSize}}|*{39}{p{\cellSize}}|*{25}{p{\cellSize}}|*{42}{p{\cellSize}}|*{9}{p{\cellSize}}|}
        \hline
        $\mcal{M}$       & \multicolumn{1}{c|}{\#} & \multicolumn{25}{c|}{Set 1} & \multicolumn{39}{c|}{Set 2} & \multicolumn{25}{c|}{Set 3} & \multicolumn{42}{c|}{Set 4} & \multicolumn{9}{c|}{Set 5} \\ \hline
        \opSource        & 203  & \f&\p&\p&&&\p&\p&\f&\f&\p&\p&\f&\f&\f&\p&\p&\p&\p&\f&\f&\f&\f&\f&\p&\p&\p&\p&\p&\p&\p&\f&\f&\p&\p&\p&\p&\p&\p&\p&\p&&\p&\p&\p&\p&\p&\p&\p&\p&\p&\p&\f&\p&\p&\p&\f&\p&\p&\p&\p&\p&\p&\p&\p&\p&\p&\p&\p&\p&\p&\p&\p&\p&\p&\p&\p&\p&\p&&\f&\f&\p&\p&\f&\f&\p&\f&\p&\p&\p&\p&\p&\p&\p&\p&\p&\p&\p&\p&&\p&\p&\p&\p&\p&\p&\p&\p&\p&\p&\p&\p&\f&\p&&\f&\f&\f&\p&\p&\p&\f&\f&\f&\p&\p&\f&\p&\p&\p&\p&\p&\p&\p&\f&\p&\f&\p&\f&\p\\ \cline{1-2}
        \opTarget        & 333  & \p&\p&\p&&\f&\p&\p&\f&\f&\p&\p&\p&\p&\p&\p&\p&\p&\p&\f&\p&\p&\p&\p&\p&\p&\p&\p&\p&\p&\p&\p&\p&\p&\p&\p&\p&\p&\p&\p&\p&\p&\p&\p&\p&\p&\p&\p&\p&\p&\p&\p&\p&\p&\p&\p&\p&\p&\p&\p&\p&\p&\p&\p&\f&\p&\p&\p&\p&\p&\p&\p&\p&\p&\p&\p&\p&\p&\p&&\p&\p&\p&\p&\f&\p&\p&\p&\p&\f&\p&\p&\p&\p&\p&\p&\p&\p&\p&\p&\p&\p&\p&\p&\p&\p&\p&\p&\p&\p&\p&\p&\p&\p&\p&&\p&\p&\p&\p&\p&\p&\f&\p&\p&\p&\p&\p&\p&\p&\f&\f&\p&\p&\p&\p&\p&\p&\f&\p&\p\\ \cline{1-2}
        \opOut           & 143  & \p&\p&\p&&\f&\p&\p&\f&\f&\p&\p&\f&\f&\f&\p&\p&\p&\p&\f&\f&\f&\f&\f&\p&\p&\p&\p&\p&\p&\p&\f&\f&\p&\p&\p&\p&\p&\f&\p&\p&&\p&\p&\p&&\p&\p&\p&\p&\p&\p&\f&\p&\p&\p&\f&\p&\p&\p&\p&\p&\p&\p&&\p&\p&\p&\p&\p&\p&\p&\p&\p&\p&\p&\p&\p&\p&&\f&\p&\p&\p&\f&\f&\p&\f&\p&&\p&\p&\p&\p&\p&\p&\f&\p&\p&\p&&\p&\p&\p&\p&\p&\p&\p&&\p&\p&\p&\p&\f&\p&&\f&\f&\p&\p&\p&\p&\f&\f&\f&\p&\p&\f&\p&\p&&&\p&\p&\p&\f&\p&\f&\p&\f&\p\\ \cline{1-2}
        \opInv           & 7 & \f&\f&\p&&&\p&\p&\f&\f&\f&\f&\f&\p&\p&\f&\f&\f&\f&\f&\f&\f&\f&\f&\p&\p&\p&\p&\f&\f&\f&\p&\f&\p&\p&\p&&\p&&\p&&&&\p&&&&\p&\p&\p&\f&\f&\p&\f&\f&\f&\f&&&&\p&\p&\p&\p&&\p&\p&&\p&\p&\p&&&\f&\f&&&\p&\f&&&\f&\f&\f&\f&\p&&\f&\p&&\p&\f&&\p&&\p&&\p&\p&&&&&\f&\p&\f&\p&&&&&\p&\p&\f&\f&&&\p&\f&\f&\f&\f&\f&\f&\p&&&\f&\p&\p&&&\p&\p&\f&\p&\f&\p&\f&\p&\p\\ \cline{1-2}
        \opSink          & 26  & \f&\p&\p&&\f&\p&\p&\f&\f&\f&\p&\f&\p&\p&\p&\p&\p&\p&\f&\p&\p&\p&\p&\p&\p&\p&\p&\p&\p&\p&\p&\p&\p&\p&\p&\p&\p&\p&\p&\p&\p&\p&\p&\p&\p&\p&\p&\p&\p&\p&\p&\p&\p&\p&\p&\p&\p&\p&\p&\p&\p&\p&\p&\p&\p&\p&\p&\p&\p&\p&\p&\p&\p&\p&\p&\p&\p&\p&&\f&\f&\p&\p&\f&\p&\p&\p&\p&\p&\p&\p&\p&\p&\p&\p&\p&\p&\p&\p&\p&\p&\p&\p&\p&\p&\p&\p&\p&\p&\p&\p&\p&\p&\p&&\f&\p&\f&\p&\p&\p&\f&\p&\p&\p&\p&\p&\p&\p&\p&\p&\p&\p&\p&\p&\f&\p&\f&\p&\p\\ \cline{1-2}
        \opClockReset    & 7  & \f&\f&\f&&&\p&\p&\f&\f&\f&\p&\f&\f&\f&\p&\p&\p&\p&\f&\f&\f&\f&\f&\p&\p&\f&\p&\p&\p&\p&\f&\f&\p&\f&\f&&\f&&\f&&&&\f&\p&&&\p&\p&\p&\p&\p&\f&\p&\p&\p&\f&\p&\p&\p&\p&\p&\p&\p&&\f&\f&&\f&\f&\f&&&\f&\f&&&\p&\p&&\f&\f&\p&\p&\f&\f&\p&\f&\p&&\f&\p&&\f&&\f&&\f&\f&&&&&\f&\f&\f&\f&\p&&&&\p&\p&\f&\p&&\f&\f&\f&\p&\p&\p&\f&\f&\f&\p&\p&\f&\p&\p&&&\f&\p&\p&\f&\f&\f&\f&\f&\p\\ \cline{1-2}
        \opIn            & 30  & \f&\p&\p&&&\p&\p&\f&\f&\f&\p&\f&\f&\f&\p&\p&\p&\p&\f&\f&\f&\f&\f&\p&\p&\p&\p&\p&\p&\p&\f&\f&\p&\p&\p&\p&\p&\p&\p&\p&&\p&\p&\p&\p&\p&\p&\p&\p&\p&\p&\f&\p&\p&\p&\f&\p&\p&\p&\p&\p&\p&\p&&\p&\p&\p&\p&\p&\p&\p&\p&\p&\p&\p&\p&\p&\p&&\f&\f&\p&\p&\f&\f&\p&\f&\p&&\p&\p&\p&\p&\p&\p&\p&\p&\p&\p&&\p&\p&\p&\p&\p&\p&\p&\p&\p&\p&\p&\p&\f&\p&&\f&\f&\f&\p&\p&\p&\f&\f&\f&\p&\p&\f&\p&\p&&&\p&\p&\p&\f&\f&\f&\f&\f&\p\\ \cline{1-2}
        \opGuardConstant & 10  & \f&\f&\f&\p&\f&\p&\p&\f&\f&\f&\f&\f&\f&\f&\p&\p&\p&\p&\f&\f&\f&\f&\f&\f&\f&\f&\p&\f&\f&\p&\f&\f&\f&\f&\f&&\f&&\f&&&&\f&\p&&&\p&\p&\p&\f&\f&\f&\p&\p&\p&\f&\p&\p&\p&\f&\f&\f&\f&&\f&\f&&\f&\f&\f&&&\f&\f&&&\p&\f&\p&\f&\f&\f&\p&\f&\f&\p&\f&\f&&\f&\p&&\f&&\f&&\f&\f&&&&&\f&\f&\f&\f&\p&&&&\p&\p&\f&\f&\p&\f&\f&\f&\f&\p&\p&\f&\f&\f&\p&\p&\f&\f&\f&&&\f&\p&\f&\f&\f&\f&\f&\f&\f\\ \cline{1-2}
        \opGuardOpClocks & 8  & \p&\p&\f&&\p&&&\f&\f&\f&\f&\p&\f&\f&\p&\p&\p&\p&\p&\f&\f&\f&\f&\f&\f&\f&&\f&\f&\p&\f&\f&\f&\f&\f&&\f&&\f&&&&\f&\f&&&\f&\f&&\f&\f&\f&\p&\p&\p&\f&\p&\p&\p&\f&\f&\f&\f&&\f&\f&&\f&\f&\f&&&\p&\p&&&\f&\f&&\p&\p&\f&\p&\p&\f&\p&\f&\f&&\f&\p&&\f&&\f&&\f&\f&&&&&\p&\f&\p&\f&\f&&&&\f&&\f&\f&&\p&\f&\p&\f&\p&\p&\p&\f&\f&\p&\p&\f&\f&\f&&&\f&&\f&\f&\f&\f&\f&\f&\f\\ \cline{1-2}
        \opGuardOpVars   & 4 & \f&\f&\f&\p&&\p&\p&&&&&&&&&&&&&&&&&&&\f&\p&&&&&&&\f&\f&&\f&&\f&&&&\f&&&&\p&\p&\p&&&&&&&&&&&&&&&&\f&\f&&\f&\f&\f&&&\f&\f&&&\p&&\p&&&&&&&&&&&\f&&&\f&&\f&&\f&\f&&&&&\f&\f&\f&\f&&&&&\p&\p&&&\p&&&&&&&&&&&&&&&&&\f&\p&&&&&&&\\ \cline{1-2}
        \opVarUpdate     & 1  & \f&\f&\f&&&\p&\p&&&&&&&&&&&&&&&&&&&\f&\p&&&&&&&\f&\f&&\f&&\f&&&&\f&&&&\p&\p&\p&&&&&&&&&&&&&&&&\f&\f&&\f&\f&\f&&&\f&\f&&&\p&&&&&&&&&&&&&\f&&&\f&&\f&&\f&\f&&&&&\f&\f&\f&\f&&&&&\p&\p&&&&&&&&&&&&&&&&&&&&\f&\p&&&&&&&\\ \hline
        Total            & 772 & \p&\p&\p&\p&\p&\p&\p&\f&\f&\p&\p&\p&\p&\p&\p&\p&\p&\p&\p&\p&\p&\p&\p&\p&\p&\p&\p&\p&\p&\p&\p&\p&\p&\p&\p&\p&\p&\p&\p&\p&\p&\p&\p&\p&\p&\p&\p&\p&\p&\p&\p&\p&\p&\p&\p&\p&\p&\p&\p&\p&\p&\p&\p&\p&\p&\p&\p&\p&\p&\p&\p&\p&\p&\p&\p&\p&\p&\p&\p&\p&\p&\p&\p&\p&\p&\p&\p&\p&\p&\p&\p&\p&\p&\p&\p&\p&\p&\p&\p&\p&\p&\p&\p&\p&\p&\p&\p&\p&\p&\p&\p&\p&\p&\p&\p&\p&\p&\p&\p&\p&\p&\p&\p&\p&\p&\p&\p&\p&\p&\p&\p&\p&\p&\p&\p&\p&\p&\p&\p&\p\\ \hline
    \end{tabular}
    
    \newcommand{\pCell}{{%
        \begin{tabular}{p{\cellSize}} \p \\ \end{tabular}%
    }}
    \newcommand{\fCell}{{%
        \begin{tabular}{p{\cellSize}} \f \\ \end{tabular}%
    }}
    \newcommand{\vLine}{{%
        \begin{tabular}{l|l} \\ \end{tabular}%
    }}
    \caption{Results of testing the faulty systems. We denote by \# the number of generated test-cases. For each combination of operator and faulty system we denote if the corresponding tests resolved in at least one {\color{priFailColour}primary fail} (\ \protect\pCell\ ). Otherwise, we denote if they resolved in at least one {\color{norFailColour}other fail} (\ \protect\fCell\ ). Each PIT operator creates a set of faulty systems that we divide with vertical lines (\ \protect\vLine\ ).}
    \label{tab:evaluationMutationTesting}
\end{table}

We use the \pname extension to test the systems generated with PIT. In order to speed up test execution, we test using simulated time. \pname reports no fails for 47 of the systems. Through inspection of these systems, we found that they are all equivalent to the original car alarm system. Another 91 systems crashed while testing. When an \gls{SUT} crashes, \pname provides the stack trace, which allows developers to locate the fault. However, it does not provide us with an output with which we can determine the verdict. This leaves us with 140 failing systems that do not crash. We denote these as \emph{faulty systems}. The results from testing the faulty systems can be seen in \cref{tab:evaluationMutationTesting}.

We observe that every faulty system has a failed test-case. This shows that \emph{\pname detected all the generated faults}. In three systems (\eg number 4 in set 1) \opGuardConstant and \opGuardOpVars are the only ones that could detect the fault. These operators are new and defined in this paper. This shows that the new operators improve the ability to detect faults.

As mentioned in \cref{sec:generation} \pname includes primary fails. For three systems (\eg number 5 in set 1) \opGuardOpClocks is the only one to achieve a primary fail. This operator is new and defined in this paper. This shows that the new operators improve the ability to locate faults. 

Two faulty systems (\eg number 8 in set 1) have no primary fails. Both change an output to one not defined in the test model. In order to generate primarily failed test-cases for such faults, their mutants would have to guess these non-defined outputs. As there is an infinite number of non-defined outputs, it is in practice impossible to guarantee a primary fail for every such fault.

\section{User Interface} \label{sec:ui}

\begin{figure}[tb]
    \centering
    \includegraphics[width=\textwidth]{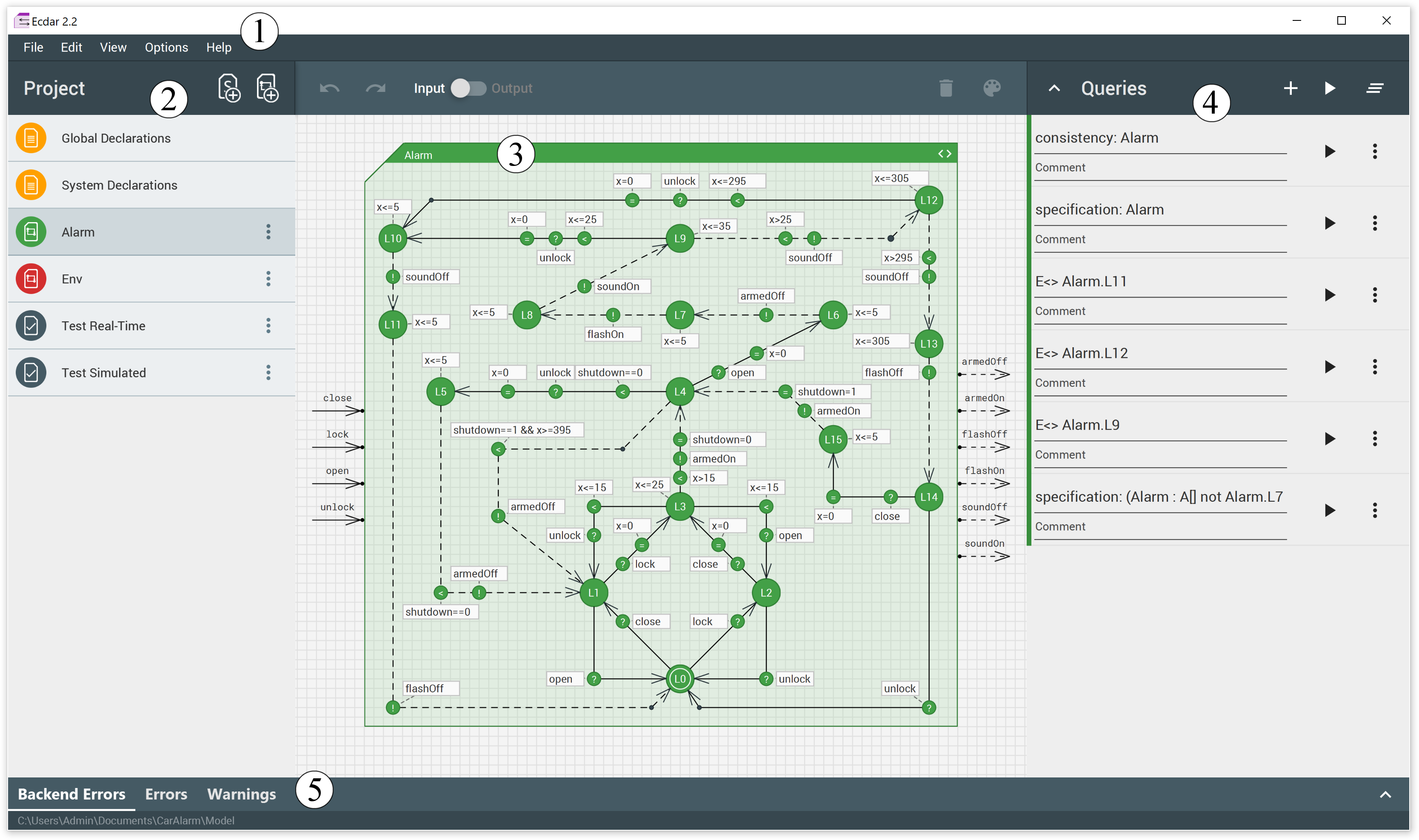}
    \caption{A screenshot of the \pname UI overlayed with numbered circles.}
    \label{fig:ecdarCanvas}
\end{figure}

\begin{figure}[tb]
    \centering
    \begin{subfigure}[b]{0.50\textwidth}
        \centering\includegraphics[width=.95\textwidth]{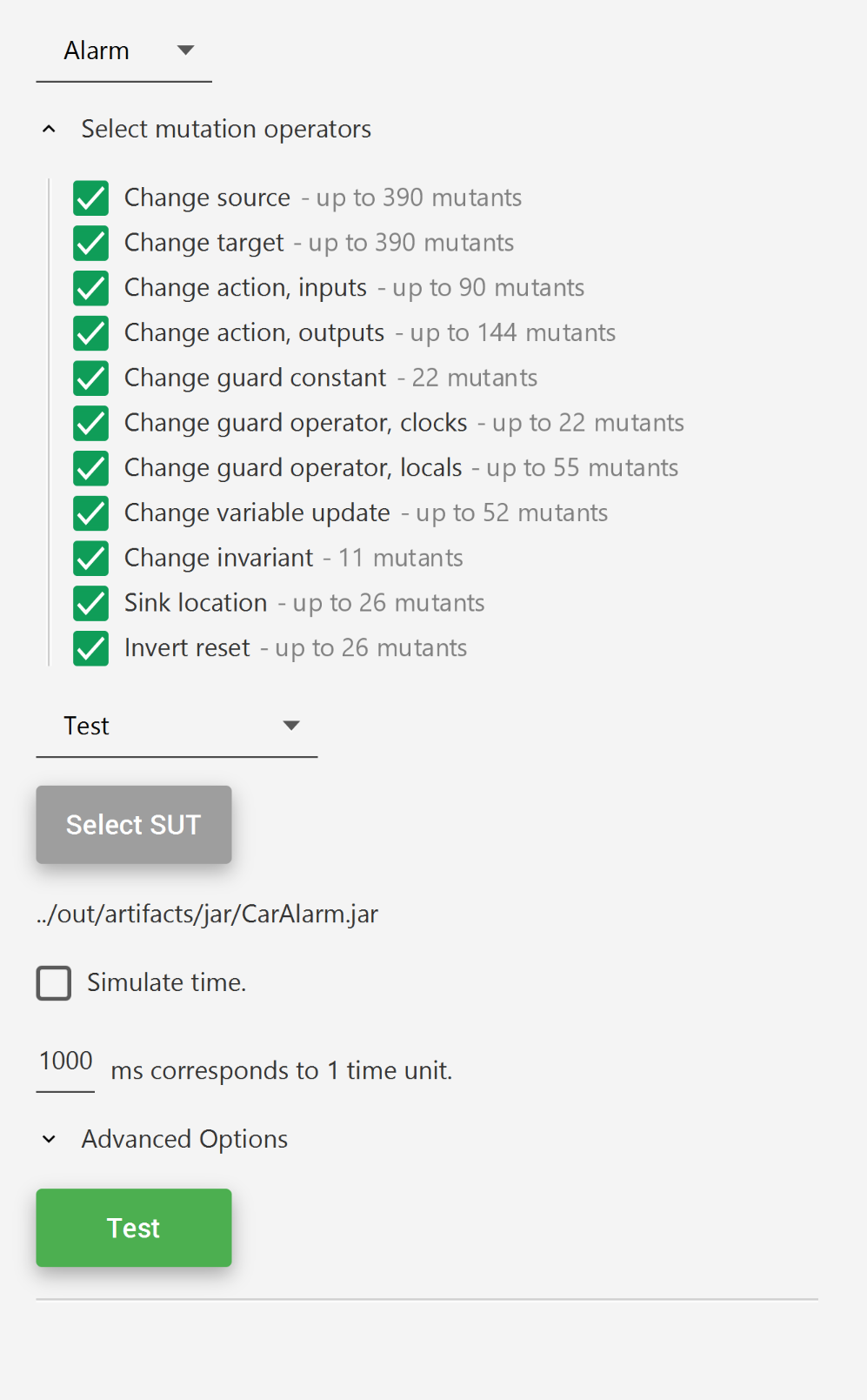}
        \caption{\label{fig:testPlan}A test plan.}
    \end{subfigure}%
    \begin{subfigure}[b]{0.50\textwidth}
        \centering\includegraphics[width=.95\textwidth]{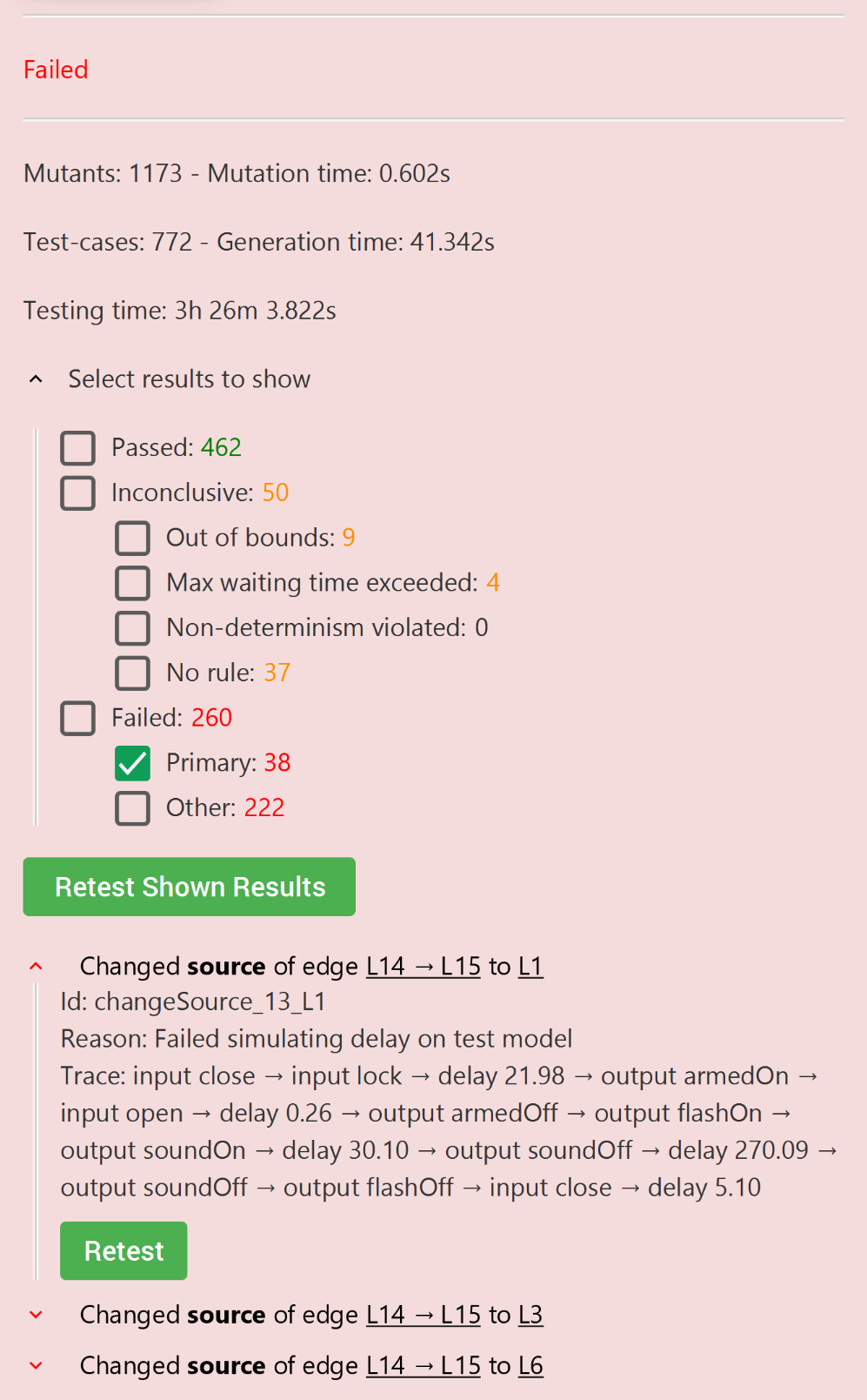}
        \caption{\label{fig:testResults}Results of a failed test run.}
    \end{subfigure}
    \caption{Partial screenshots of an \pname UI.}
    \label{fig:testScreens}
\end{figure}

The \ecdar extension is implemented as a program that works as an IDE. \cref{fig:ecdarCanvas} shows a screenshot of \pname, where:

\begin{enumerate}
    \item[\circled{1}] from the menu bar, we can create new documents and change options;
    \item[\circled{2}] from the project pane, we can manage the declarations of the system, the \glspl{TIOA}, and the test plans;
    \item[\circled{3}] on the canvas, we can edit \glspl{TIOA}. For instance, the \glspl{TIOA} in \cref{fig:tioaExample,fig:alarm} are modelled this way;
    \item[\circled{4}] on the query pane, we can use the \ecdar engine to verify \gls{TCTL} properties; and
    \item[\circled{5}] on the error pane, errors and warnings automatically appear.
\end{enumerate}

From the menu bar \circled{1}, we can create a new test plan. In order to test, we simply need to choose our test model, select the path to our \gls{SUT} or a program interfacing it, and (if testing using real-time) define what 1 time unit in the model corresponds to in real-time (see \cref{fig:testPlan}). The current version of \pname only works with an \gls{SUT} that is a JAR file or interfaced by one.

Testers can adjust a test plan by changing:
\begin{itemize}
    \item what mutation operators to test with,
    \item whether to test using simulated time,
    \item the maximum number of concurrent threads used for generation,
    \item the maximum number of concurrent instances of the \gls{SUT},
    \item the maximum wait time, and 
    \item the step bound.
\end{itemize}

For each executed test-case, \pname displays a description of the mutant, the id of the test-case, the verdict, reason for the verdict, and the trace performed by the \gls{SUT} (see \cref{fig:testResults}). The tester should use this information to locate faults.

\pname supports retesting selected test-cases without redoing mutation or test-case generation.
It can also export mutants for custom use.
The extension and sample projects (including the car alarm presented in \cref{sec:caseStudy}) are licensed under MIT\footnote{\url{https://opensource.org/licenses/MIT}} and are available at \url{http://ulrik.blog.aau.dk/ecdar/ecdar-2-2/}.

\section{Conclusion}\label{sec:conclusion}

\glsresetall

In this paper, we present an extension of \ecdar that integrates conformance testing into the tool in order to improve productiveness and reliability. It uses \gls{MBMT} that -- contrary to similar approaches -- is fault-based, proving the absence of certain types of faults. It also generates test-cases solely based on the test model. Thus, the tester does not need to provide any other constructs unlike what is required for some other methods.

\pname mutates the test model through 11 mutation operators and uses the \ecdar engine to generate strategies that we use for test-case generation. The tool then executes the test-cases using either real-time or simulated time. Testing using real-time enables testing of physical systems. Testing using simulated time allows for a significant speed-up. To further speed up testing, we parallelise test-case generation and test execution.

\pname is an open-source IDE using the \ecdar engine. It can model \glspl{TIOA}, verify them, and test a system based on a \gls{TIOA}.

We conduct a case study using the independent mutation testing tool PIT to generate 140 faulty systems. \emph{Testing with \pname detected all faults}. 
We introduce new mutation operators that improve the ability to detect and locate faults.

\paragraph{Future Work}

\ecdar can combine models with conjunction, composition, and quotient. It does, however, only support testing of one uncombined model at a time. This makes it tedious to test integrated systems and makes it unusable for integration testing. Future work could include support for testing integrated systems.
The challenge is to simulate combined automata. We could take inspiration from \uppaalTron \cite{Hessel2008TronTool}, as it solves a similar problem; it does not know the state of the \gls{SUT}, so it keeps track of multiple potential states.
When testing integrated systems, we could check refinements using compositional verification \cite{David2012CompVeriWithEcdar} in order to speed up test-case generation.

We presented in \cref{sec:caseStudy} a small case study to demonstrate \gls{MBMT} with \pname. However, future work could include an industrial case study.

\pname only supports testing using first-order mutants. Future work could include using higher-order mutants to detect faults not detectable by first-order mutants.
Future work could also include reducing generation and execution time. An approach is to prioritise test-cases so that we only generate and execute test-cases with a higher probability of detecting a fault. Alternatively, Lorber \etal \cite{Lorber2018viaModelChecking} and Devroey \etal \cite{Devroey2016FeaturedMutantModel} propose other approaches for speeding up, as presented in \cref{sec:related}.

Finally, future work could include adding a command-line interface to test with \pname. We can, for instance, use this to conformance test during continuous integration.

\ifdefined\GandALF
\else
    \paragraph{Acknowledgements} We thank Ulrik Nyman for supervision, and Florian Lorber and Ulrik for fruitful discussions throughout the project period.
\fi

\bibliographystyle{eptcs}
\bibliography{References/P9,References/MasterProject}
\end{document}